\documentclass[%
12pt,
reprint,
superscriptaddress,
preprintnumbers,
amsmath,amssymb,
aps,
pre,
]{revtex4-1}

\usepackage{isomath}
\usepackage{amsmath}
\usepackage{amsbsy}
\usepackage{amssymb}
\usepackage{amscd}
\usepackage{amsfonts}
\usepackage{stmaryrd}
\usepackage{units}
\usepackage{euscript}
\usepackage[utf8]{inputenc}
\usepackage[T1]{fontenc}
\usepackage{times}
\everymath{\displaystyle}
\usepackage{exscale}
\usepackage[plain]{fancyref}
\usepackage{mathtools}
\usepackage{soul}

\newcommand*{\fancyrefapplabelprefix}{app}
\frefformat{plain}{\fancyrefapplabelprefix}{appendix~#1}
\Frefformat{plain}{\fancyrefapplabelprefix}{Appendix~#1}
\frefformat{main}{\fancyrefapplabelprefix}{appendix~#1}
\Frefformat{main}{\fancyrefapplabelprefix}{Appendix~#1}

\usepackage{graphicx}
\usepackage{boxedminipage}
\usepackage{calc}
\usepackage[usenames,dvipsnames]{xcolor}
\newcommand*{\gnuplotinput}[2][1.0]{%
	\begingroup
	\let\@gnplt@input@includegraphics=\includegraphics
	\def\includegraphics##1{\@gnplt@input@includegraphics[scale=#1]{#2}}%
	\let\@gnplt@input@picture=\picture
	\def\picture{\unitlength=#1\unitlength\relax\@gnplt@input@picture}%
	\input{#2}%
	\endgroup
}
\makeatother

\usepackage[small,compact]{titlesec}
\usepackage{setspace}
\usepackage{paralist}
\usepackage{enumitem}
\setitemize{noitemsep,topsep=0pt,parsep=0pt,partopsep=0pt}
\setenumerate{noitemsep,topsep=0pt,parsep=0pt,partopsep=0pt}
\setdescription{noitemsep,topsep=0pt,parsep=0pt,partopsep=0pt}

\usepackage{verbatim}

\usepackage[,colorlinks,urlcolor=blue,citecolor=blue]{hyperref}
\usepackage[normalem]{ulem}

\usepackage{isomath}
\usepackage{amsmath}
\usepackage{amssymb}
\usepackage{amscd}
\usepackage{amsfonts}

\newcommand{\N}{\mathbb N}

\DeclareMathOperator{\erf}{erf}

\DeclareMathOperator{\grad}{grad}

\DeclareMathOperator{\curl}{curl}

\newcommand{\bfe}{{\mathbold e}}

\newcommand{\bfn}{{\mathbold n}}

\newcommand{\bfp}{{\mathbold p}}

\newcommand{\bfr}{{\mathbold r}}

\newcommand{\bft}{{\mathbold t}}
\newcommand{\bfu}{{\mathbold u}}
\newcommand{\bfv}{{\mathbold v}}

\newcommand{\bfx}{{\mathbold x}}
\newcommand{\bfy}{{\mathbold y}}

\newcommand{\bfD}{{\mathbold D}}
\newcommand{\bfE}{{\mathbold E}}
\newcommand{\bfF}{{\mathbold F}}

\newcommand{\bfN}{{\mathbold N}}

\newcommand{\bfP}{{\mathbold P}}

\DeclareMathOperator{\spn}{span}
\DeclareMathOperator{\diag}{diag}

\newcommand{\vol}{V}
\newcommand{\W}{\mathcal{W}^*} 
\newcommand{\F}{\bfF} 
\newcommand{\tmag}{t_0}
\newcommand{\tvec}{\mathbf{\tmag}}
\newcommand{\tnodim}{\tilde{t}_0}
\newcommand{\stressij}{\sigma}
\newcommand{\stress}{\boldsymbol{\stressij}} 
\newcommand{\mech}{\stress_{\text{mech}}}
\newcommand{\maxwellij}{\Sigma}
\newcommand{\maxwell}{\mathbf{\maxwellij}} 
\newcommand{\pos}{\mathbf{x}} 

\DeclareMathOperator{\Grad}{Grad}
\DeclareMathOperator{\Curl}{Curl}
\DeclareMathOperator{\Div}{Div}
\newcommand{\Rvec}{\tilde{\mathbf{r}}}
\newcommand{\chaindensity}{N}
\newcommand{\pstch}[1]{\lambda_{#1}}
\newcommand{\cbody}{\Omega}

\newcommand{\cbndy}{\partial \cbody}

\newcommand{\avg}[1]{\left\langle #1 \right\rangle_{\Rvec}}
\newcommand{\epot}{\varphi}

\newcommand{\diverge}[1]{\text{ div} #1}

\newcommand{\vperm}{\epsilon_0}

\newcommand{\pstche}{\pstch{\emag}}

\newcommand{\delen}{\ell} 
\newcommand{\deth}{t} 
\newcommand{\susceptibility}{\chi}

\newcommand{\polarizationmag}{P}
\newcommand{\polarization}{\mathbf{\polarizationmag}}

\newcommand{\shearDef}{s}

\newcommand{\A}{\mathcal{F}}

\newcommand{\T}{T}

\newcommand{\kB}{k}

\newcommand{\mlen}{b}

\renewcommand{\N}{n}

\newcommand{\density}{\rho}

\newcommand{\rmag}{r}
\newcommand{\rvec}{\bfr}
\newcommand{\rz}{\rmag_3}
\newcommand{\rx}{\rmag_1}

\newcommand{\rdir}{\hat{\rvec}}

\newcommand{\stch}{\gamma}
\newcommand{\stchx}{\stch_1}
\newcommand{\stchz}{\stch_3}
\newcommand{\stchr}{\stch_r}

\newcommand{\unitsphere}{\mathbb{S}^2}

\newcommand{\um}{u}

\newcommand{\nvec}{\hat{\mathbf{n}}}

\newcommand{\sussymbol}{\chi}
\newcommand{\sus}[1]{\ifthenelse{#1 < 2}{\sussymbol_{\parallel}}{\sussymbol_{\perp}}}
\newcommand{\sustens}{\boldsymbol{\chi}}
\newcommand{\dsus}{\Delta \sussymbol}

\newcommand{\emag}{E}
\newcommand{\efield}{\bfE}
\newcommand{\edir}{\hat{\efield}}

\newcommand{\dipolemag}{\mu}
\newcommand{\dipole}{\boldsymbol{\dipolemag}}

\newcommand{\chainpolar}{\bfp}

\newcommand{\unodim}{\kappa}
\newcommand{\uxnodim}{\unodim_1}
\newcommand{\uznodim}{\unodim_{3}}

\newcommand{\uOnodim}{\unodim_{\perp}}
\newcommand{\uslnodim}{\unodim}

\newcommand{\lct}{\varepsilon}

\newcommand{\im}{{i\mkern1mu}}

\newcommand{\erfi}{\mathrm{erfi}}

\newcommand{\dirac}{\delta}

\newcommand{\multmag}{\tau}
\newcommand{\mults}{\boldsymbol{\multmag}}

\newcommand{\zmult}{\lambda}
\newcommand{\xmult}{\alpha}

\newcommand{\C}{C}

\newcommand*\df{\mathop{}\!\mathrm{d}}

\newcommand{\iden}{\mathbf{I}}

\newcommand{\Lang}{\mathcal{L}}
\newcommand{\Langinv}{\Lang^{-1}}
\newcommand{\Langinvs}{\Langinv\left(\stch\right)}

\newcommand{\csch}{\mbox{ csch }}

\newcommand{\eone}{\hat{\bfe}_1}
\newcommand{\etwo}{\hat{\bfe}_2}
\newcommand{\ethree}{\hat{\bfe}_3}

\newcommand{\densitykg}{\density_{KG}}

\newcommand{\Asm}{\A_{s\multmag}}
\newcommand{\Akg}{\A_{KG}}

\newcommand{\order}[1]{\mathcal{O}\left(#1\right)}

\newcommand{\lag}{L}
\newcommand{\mult}{\zeta}
\newcommand{\pbE}{\tilde{\bfE}}
\newcommand{\pbD}{\tilde{\bfD}}
\newcommand{\pbP}{\tilde{\bfP}}
\newcommand{\free}{\Psi}
\newcommand{\Wm}{\mathcal{W}_m}
\newcommand{\We}{\mathcal{W}_e}

\newcommand{\pwrt}[2]{\frac{\partial #1}{\partial #2}}

\newcommand{\pmStress}{\tilde{\boldsymbol{\Sigma}}}

\newcommand{\sust}{\boldsymbol{\chi}}
\newcommand{\tStress}{\mathbf{T}}

\newcommand{\sustensexpr}{
	\sus{1} \nvec \otimes \nvec + \sus{2} \left(\iden - \nvec \otimes \nvec\right)
}

\newcommand{\intoverSns}[1]{\int_{\unitsphere} \df{A} \mbox{ } #1}

\newcommand{\zmultzero}{\Langinv\left(\stch\right)}

\newcommand{\erfw}{\erf \left(\sqrt{\uslnodim}\right)}
\newcommand{\Csl}{\frac{\N \sqrt{\uslnodim}}{2 \pi^{3/2} \erfw}}
\newcommand{\zmultsl}{\frac{2\sqrt{\pi} \stchz \uslnodim e^{\uslnodim} \erfw}{\sqrt{\pi} e^{\uslnodim} \erfw - 2\sqrt{\uslnodim}}}
\newcommand{\xmultsl}{\frac{4\sqrt{\pi} \stchx \uslnodim e^{\uslnodim} \erfw}{\sqrt{\pi} \left(2\uslnodim-1\right)  e^{\uslnodim} \erfw + 2\sqrt{\uslnodim}}}

\renewcommand{\hl}[1]{}
\graphicspath{ {figs/} }

\begin{document}


\preprint{To appear in Phys. Rev. E (\url{https://doi.org/10.1103/PhysRevE.103.042504})}

\title{\Large{Nonlinear Statistical Mechanics Drives Intrinsic Electrostriction and Volumetric Torque in Polymer Networks}}

\author{Matthew Grasinger}
    \email{matthew.grasinger.ctr@afresearchlab.com}
    \affiliation{Pittsburgh Quantum Institute, University of Pittsburgh}
    \affiliation{Department of Civil and Environmental Engineering, Carnegie Mellon University}
    \affiliation{UES, Inc.}
    \affiliation{Materials and Manufacturing Directorate, Air Force Research Laboratory}

\author{Carmel Majidi}
    \affiliation{Department of Civil and Environmental Engineering, Carnegie Mellon University}
    \affiliation{Department of Mechanical Engineering, Carnegie Mellon University}
    \affiliation{Department of Materials Science and Engineering, Carnegie Mellon University}

\author{Kaushik Dayal}
    \affiliation{Pittsburgh Quantum Institute, University of Pittsburgh}
    \affiliation{Department of Civil and Environmental Engineering, Carnegie Mellon University}
    \affiliation{Department of Materials Science and Engineering, Carnegie Mellon University}
    \affiliation{Center for Nonlinear Analysis, Department of Mathematical Sciences, Carnegie Mellon University}

\date{\today}

\begin{abstract}
    Statistical mechanics is an important tool for understanding polymer electroelasticity because the elasticity of polymers is primarily due to entropy.
    However, a common approach for the statistical mechanics of polymer chains, the Gaussian chain approximation, misses key physics.
    By considering the nonlinearities of the problem, we show a strong coupling between the deformation of a polymer chain and its dielectric response; that is, its net dipole.
    When chains with this coupling are cross-linked in an elastomer network and an electric field is applied, the field breaks the symmetry of the elastomer's elastic properties, and, combined with electrostatic torque and incompressibility, leads to intrinsic electrostriction.
    Conversely, deformation can break the symmetry of the dielectric response leading to volumetric torque (i.e., a couple stress or torque per unit volume) and asymmetric actuation.
    Both phenomena have important implications for designing high-efficiency soft actuators and soft electroactive materials; and the presence of mechanisms for volumetric torque, in particular, can be used to develop higher degree of freedom actuators and to achieve bioinspired locomotion.
\end{abstract}

\maketitle

\section{Introduction}

\begin{figure} 
  \includegraphics[clip, trim=1.0cm 1.0cm 1.0cm 1.0cm, width=\linewidth]{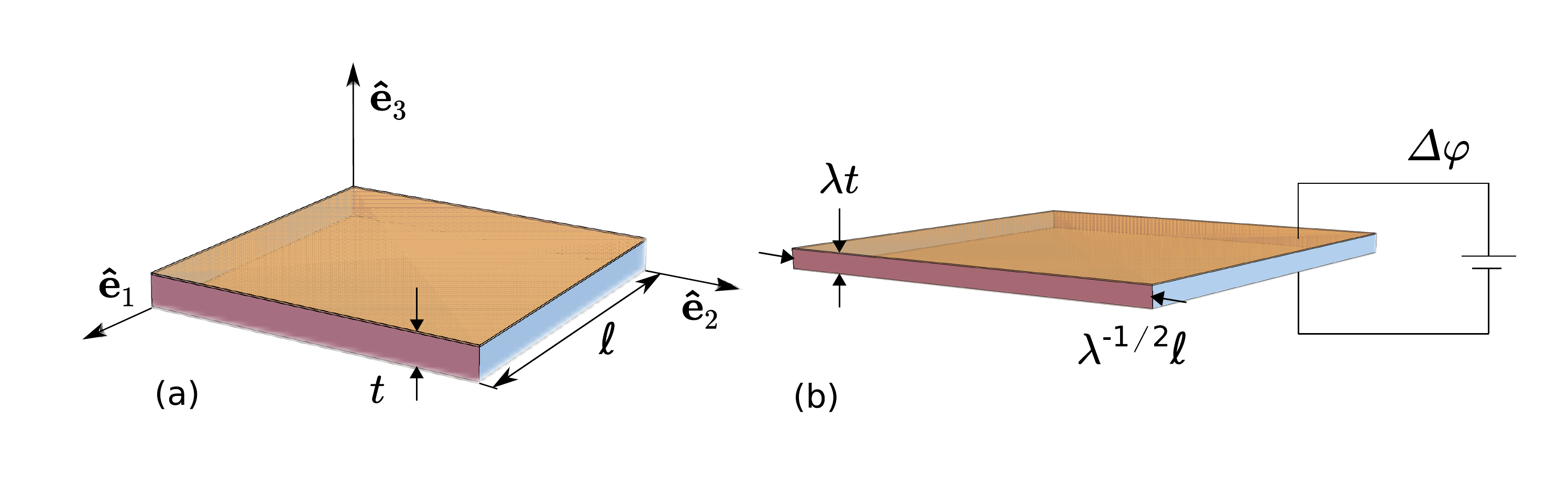}
	\caption{Dielectric elastomer actuator with dimensions of length $\delen$ in the $\eone$ and $\etwo$ directions, and thickness $\deth$ in the $\ethree$ direction, where $\deth \ll \delen$: (a) undeformed configuration; (b) a voltage difference is applied across the electrodes on the top and bottom surfaces and, as a result, the actuator contracts across its thickness and expands in the plane orthogonal.}
	\label{fig:DEA}
\end{figure}

Dielectric elastomers (DEs) respond to electric fields with large deformations and can be used to convert between electrical and mechanical energy.
Therefore, DEs are promising for applications in energy harvesting, biomedical devices, and soft, biologically-inspired robotics~\cite{bar-cohen2001electroactive,carpi2011electroactive,kim2007electroactive,huang2012giant,majidi2014soft,bartlett2016stretchable,ware2016localized,erol2019microstructure,castaneda2011homogenization,galipeau2013finite,liu2013giant,deng2014electrets,zhang2017nonlinear}.
The quintessential example of a DE actuator (DEA) is a thin DE film sandwiched between two compliant electrodes.
When a voltage difference is applied across the electrodes, the DE polarizes and compresses across its thickness and, because of the Poisson effect, expands in the plane of the electrodes (see \Fref{fig:DEA}).
Previously, this deformation has been understood as entirely a consequence of the Coulomb attraction between the electrodes which squeezes together the top and bottom of the film~\cite{pelrine2000high,wissler2007mechanical,kofod2008static,kollosche2012complex,pelrine2001dielectric}.

Elastomers consist of thermally fluctuating polymer chains linked together at various junctions (i.e. cross-links).
Since these thermal fluctuations are significant, statistical mechanics has provided much of our current understanding of polymer elasticity~\cite{treloar1975physics,flory1944network}.
In DEs, the polymers polarize in the presence of an electric field; and since the dipoles on the monomers prefer to be aligned with the field, their electrostatic interaction affects how the chains fluctuate (and vice versa, chain deformation affects the electrostatics).
To consistently capture the electroelasticity of DEs, one must consider the electrostatics as part of the statistical mechanics formulation.

In this work, we begin with a mechanism for polarization at the monomer-scale and then use nonlinear (i.e. non-Gaussian) statistical mechanics and orientational averaging over chains (i.e. network modeling) to derive the free energy functional of DEs, which implies a deformation dependent susceptibility.
The deformation dependence is such that the polarization response may be anisotropic, which leads to a rich interplay between deformation and polarization.
We show electromechanical couplings at the continuum-scale that cannot be captured through an isotropic polarization response and/or the Coulomb effects alone.
Specifically, 
\begin{inparaenum}[1)] 
    \item we see significant \emph{intrinsic} electrostriction, in that there is significant deformation beyond that expected by simple Coulombic attraction of the electrodes; and 
    \item discover an asymmetric shear-mode of electromechanical actuation--driven by volumetric torque (by ``volumetric torque'', we mean a torque per unit volume~\footnote{In the literature, this is also sometimes referred to as ``body couples'' or ``couple stresses'' (see \S 4.3 of~\cite{tadmor2012continuum} or \S 2.3.2 of~\cite{tadmor2011modeling}). 
        Volumetric torques often arise as conjugates to the orientational degrees of freedom that are required to describe materials such as liquid crystals \cite{de1993physics}.})--which is a consequence of the deformation breaking the symmetry of the polarization response. 
\end{inparaenum}
Both phenomena have important implications for designing high-efficiency soft actuators and soft electroactive materials.
The presence of mechanisms for volumetric torque, in particular, can be used to develop high-degree of freedom actuators with shear-modes of coupling, bending-couplings, shape morphing, etc.
Such actuators could better mimic the range of motion found in biological organisms and lead to simpler, more robust soft robots.

In previous work, it has been typical to decompose the stress into mechanical stress, $\mech$, and electrical (or Maxwell) stress, $\maxwell$ ~\cite{tutcuoglu2014energy,he2009dielectric,henann2013modeling,huang2012giant,bozlar2012dielectric,kollosche2012complex,kofod2008static,wissler2007mechanical,zhao2007electromechanical,pelrine2001dielectric,zhang2016method,li2016voltage,zhao2010theory,zurlo2017catastrophic,zhang2017nonlinear}.
The stress decomposition is written as:
\begin{equation}
    \stress = \mech + \maxwell = \mech + \polarization\left(\efield, \F, ...\right) \otimes \efield + \frac{\vperm}{2} \left|\efield\right|^2 \mathbf{I}
\end{equation}
where, $\polarization$ is the polarization, $\vperm$ is the vacuum permittivity, and by the notation $\polarization = \polarization\left(\efield, \F, ...\right)$ we mean to emphasize that the polarization response may be a function of the electric field, $\efield$, the deformation gradient, $\F$, and other state variables.
Citing the isotropic nature of elastomers, it is often assumed: $\polarization\left(\efield\right) = \vperm \susceptibility \efield$ where $\susceptibility$ is a scalar representing the susceptibility of the material~\cite{tutcuoglu2014energy,he2009dielectric,henann2013modeling,huang2012giant,bozlar2012dielectric,kollosche2012complex,kofod2008static,wissler2007mechanical,zhao2007electromechanical,pelrine2001dielectric,zhang2016method,li2016voltage,zhao2010theory,zurlo2017catastrophic,zhang2017nonlinear}.
However, it has been observed at the macroscopic scale (both theoretically and experimentally) that if the susceptibility is a function of deformation, then an additional stress develops in the dielectric~\cite{zhao2008electrostriction,suo2010theory,cohen2016electromechanical,grasinger2020architected,zhao2010theory}.
In this work, we predict a polarization response of the form:
$
\polarization\left(\efield, \F\right) = \vperm \boldsymbol{\susceptibility}\left(\efield, \F\right) \efield,
$
where $\boldsymbol{\susceptibility}$ is a second-order tensor and the direction of the polarization can vary from being aligned with $\edir$~\footnote{This more general constitutive relationship, however, still recovers the isotropic polarization response in the absence of deformation.}.
It is necessary that $\boldsymbol{\susceptibility}$ is a function of $\F$ for capturing the intrinsic electrostriction, and the anisotropy of $\boldsymbol{\susceptibility}$ is necessary for volumetric torque.

Both the intrinsic electrostriction and shear-mode electromechanical actuations are a consequence of the strong coupling, at the macromolecular-scale, between the chain deformation and its net dipole.
In the context of statistical mechanics, we model this coupling by working in a fixed end-to-end vector ensemble.
However, we highlight that the commonly used linear approximation, the Gaussian-chain~\cite{warner2007liquid}--that is, modeling the chain as a random walk biased by the electric field--does not properly capture the coupling of interest.
Physically, biasing the random walk only leads to a one-way coupling: monomer dipoles aligning with the field causes directionality on the chain elasticity, but mechanically stretching the chain does not change the densities of monomer directions.

The remainder of the paper is structured as follows:
\begin{itemize}
  \item \Fref{sec:stress} formulates the extra contribution to the stress, at the continuum-scale, that arises when an elastic dielectric has a deformation dependent susceptibility.
  \item \Fref{sec:chain} develops the nonlinear statistical mechanics of a dielectric elastomer chain and \Fref{sec:short-comings} shows how linear approaches miss key physics.
  \item \Fref{sec:orientational} describes how orientational averaging is used to model the relationships between continuum-scale quantities (i.e. deformation and polarization) and their macromolecular counterparts (i.e. end-to-end vectors and net dipoles of chains within the polymer network).
  \item \Fref{sec:electrostriction} and \Fref{sec:shear} show the intrinsic electrostriction and shear-mode of electromechanical actuation, respectively.
\end{itemize}

\section{Stress Contribution from Deformation Dependent Susceptibility}
\label{sec:stress}
Here we use a thermodynamic formulation for elastic dielectrics to show that an additional contribution to the stress occurs when the polarization susceptibility depends on deformation gradient, $\F$.
For our illustrative system, let the body of the dielectric be denoted by $\Omega$ and $\Omega_0$ in the current and reference configurations, respectively.
A material point in the current configuration is denoted by $\bfy$ and in the reference configuration by $\bfx$.
Further, $\grad$, $\diverge$, $\curl$ are the gradient, divergence, and curl with respect to $\bfy$ and $\Grad$, $\Div$, $\Curl$ are the gradient, etc., with respect to $\bfx$.

The deformation gradient is $\F = \Grad \Phi$ where $\Phi: \bfx \mapsto \bfy, \bfy = \Phi\left(\bfx\right)$ is the deformation map.
The boundary is denoted by $\partial \Omega$ and $\partial \Omega_0$ in the current and reference configurations, respectively.
The electric field is denoted by $\bfE$ and the electric potential is $\xi$ such that $\bfE = -\grad \xi$.
We decompose contributions to the free energy density such that $\Wm = \Wm\left(\F\right)$ is a contribution to the free energy density which is purely mechanical (e.g. some hyperelastic strain energy density function, such as the neo-Hookean model) and,
$\We = \We\left(\F, \pbP\right)$ is a contribution to the free energy density which corresponds to separating bound charges when the dielectric polarizes.
Therefore the total Helmholtz free energy of the system is
\begin{equation} \label{eq:free1}
\begin{split}
  \free\left[\bfy, \pbP\right] =& \int_{\Omega_0} \df{\bfx} \left(\Wm\left(\F\right) + \We\left(\F, \pbP\right) \right) + \\
                              &\frac{\vperm}{2} \int_{\Omega} \df{\bfy} \: \emag^2 + \int_{\partial \Omega} \df{\bfy} \: \xi \bfD \cdot \bfn.
\end{split}
\end{equation}
where $\emag = |\efield|$ and $\bfn$ is the unit normal to $\partial \Omega$.

For simplicity, we assume that everywhere on the boundary either $\xi$ is specified or there are no free charges (i.e. $\bfD \cdot \bfn = 0$).
We make the following definitions:
\begin{equation} \label{eq:defs}
    \pbE = \F^T \bfE, \quad \pbD = J \F^{-1} \bfD, \quad \pbP = J \bfP,
\end{equation}
where $J = \det \F$, $\bfE$ is the electric field, $\bfP$ is the polarization, $\bfD$ is the electric displacement, and $\pbE$, $\pbP$, and $\pbD$ are their respective pullbacks to the reference configuration.
These definitions, together with Maxwell's equations, imply that
\begin{equation} \label{eq:implications}
\begin{split}
  &\Curl \pbE = 0, \quad \pbE = -\Grad \xi_0, \\
  &\pbD = \F^{-1}\left(\vperm J \F^{-T} \pbE + \pbP\right), \quad \Div \pbD = 0,
\end{split}
\end{equation}
where $\xi_0\left(\bfx\right) = \xi\left(\Phi\left(\bfx\right)\right)$~\cite{yang2017revisiting,castaneda2012finite,marshall2014atomistic,liu2013energy,liu2014energy}.
Pulling back \eqref{eq:free1} to the reference configuration, we obtain:
\begin{equation}
  \begin{split}
    \free\left[\bfy, \pbP\right] = \int_{\Omega_0} \df{\bfx} \left(\Wm\left(\F\right) + \We\left(\F, \pbP\right) + \frac{\vperm}{2} \emag^2 J\right) \\
    + \int_{\partial \Omega_0} \df{\bfx} \: \xi_0 \pbD \cdot \bfN \,,
  \end{split}
\end{equation}
where $\bfN$ is the unit normal to $\partial \Omega_0$.

Optionally, we can enforce incompressibility with a Lagrange multiplier function, $\mult = \mult\left(\bfx\right)$.
In this case, we search for conditions in which the functional:
\begin{equation} \label{eq:lag}
    \lag\left[\bfy, \pbP; \mult\right] = \free\left[\bfy, \pbP\right] - \int_{\Omega_0} \df \bfx \: \mult \left(J - 1\right),
\end{equation}
is stationary with respect to variations.

For the variation of $\lag$ with respect to $\bfy$ to vanish, we require that:
\begin{equation}
\begin{split}
    \left(\pwrt{\Wm}{\F} + \pwrt{\We}{\F} + \pmStress - J \mult \F^{-T}\right) \bfN = \mathbf{0}, \quad \text{ on } \partial \Omega_0, \\
    \Div \left(\pwrt{\Wm}{\F} + \pwrt{\We}{\F} + \pmStress - J \mult \F^{-T}\right) = \mathbf{0}, \quad \text{ on } \Omega_0.
\end{split}
\end{equation}
where $\pmStress \coloneqq \efield \otimes \pbD - \frac{\vperm}{2}\emag^2 J \F^{-T}$ is the Piola-Maxwell stress (see~\cite{yang2017revisiting,liu2014energy} for example).
We call the tensor
\begin{equation}
   \tStress \coloneqq \pwrt{\Wm}{\F} + \pwrt{\We}{\F} + \pmStress - J \mult \F^{-T}
\end{equation}
the total stress tensor and consider its various contributions.
The term $\partial \Wm / \partial \F - J \mult \F^{-T}$ is typically referred to as the ``mechanical stress''~\footnote{If the material is not incompressible, the constraint is dropped and this term simplifies to $\partial \Wm / \partial \F$.} and $\pmStress$ is considered the electrical part.
The remaining term involves $\We$, which, recall, is the contribution to the free energy density which corresponds with separating bound charges when the dielectric polarizes.
For a linear dielectric, 
\begin{equation} \label{eq:linear-dielectric}
    \We = \frac{1}{2 \vperm J} \pbP \cdot \sust^{-1} \pbP.
\end{equation}
where $\sust$ is the polarization susceptibility tensor.
If the material is incompressible, i.e. $J = 1 = \text{const}$, and $\sust$ is not a function of $\F$, as is often assumed, then the $\partial \We / \partial \F$ term vanishes.
However, in \Fref{sec:orientational} we show that, for dielectric elastomers, $\sust$ is a function of $\F$.
Thus, there is an additional contribution to the total stress which is often not accounted for:
\begin{equation}
  \frac{\partial \We}{\partial \F} = -\frac{1}{2 \vperm J} \left(\pbP \cdot \sust^{-1} \pbP\right) \F^{-T} + \frac{1}{2 \vperm J} \pbP \cdot \frac{\partial \sust^{-1}}{\partial \F} \pbP.    
\end{equation}
This contribution does not vanish even if the material is incompressible.
It is this stress contribution which leads to the intrinsic electrostriction in dielectric elastomers.

For completeness, we also note that for the variation of $\lag$ with respect to $\pbP$ to vanish, we require that:
\begin{equation}
    \pwrt{\We}{\pbP} = \efield \quad \text{ on } \Omega_0\,.
\end{equation}
Again, if we consider a linear dielectric, by \eqref{eq:linear-dielectric}
\begin{equation}
    J^{-1} \pbP = \bfP = \vperm \sust \efield,
\end{equation}
such that, when $\sust = \sust\left(\F\right)$, the material polarization response, $\bfP$, depends on the deformation.
We show in \Fref{sec:shear} that, for dielectric elastomers, not only does $\sust$ depend on $\F$, but it does so in such a way that $\bfP$ does not necessarily align with $\efield$.
When this misalignment occurs, the Maxwell stress is asymmetric and the dielectric experiences an electrostatic volumetric torque.

\newcommand{\dipoleSusceptibility}{\boldsymbol{\susceptibility}_{\dipolemag}}
\section{Statistical Mechanics of a Dielectric Elastomer Chain} \label{sec:chain}
Following the simplest classical theory \cite{treloar1975physics,kuhn1942beziehungen}, we idealize the mechanics of a polymer such that 
\begin{inparaenum}[1)] 
    \item monomers are rigid, 
    \item monomers are free to rotate about their neighboring bonds, and 
    \item excluded volume effects are neglected. 
\end{inparaenum}
This means the maximum length of the chain end-to-end vector, $\rvec$, is $\N \mlen$, where $\N$ is the number of monomers in the chain and $\mlen$ is the monomer length.
Further, in the presence of an electric field, bound charges on a monomer can be separated to form an electric dipole, $\dipole$.
Given that the dipole depends on the magnitude of the electric field and the orientation of the monomer, $\nvec$, relative to the direction of the electric field, we use a simple anisotropic form \cite{stockmayer1967dielectric,cohen2016electroelasticity}:
\begin{equation}
  \begin{split}
    \dipole\left(\nvec, \efield\right) &= \vperm \dipoleSusceptibility \efield \\ &= \vperm \left[\sustensexpr\right] \efield 
  \end{split}
\end{equation}
where $\dipoleSusceptibility$ is the dipole susceptibility tensor, $\sus{1}$ and $\sus{2}$ are the dipole susceptibility along $\nvec$ and the susceptibility in plane orthogonal to $\nvec$, respectively, $\efield$ is the local electric field, and $\vperm$ is the vacuum permittivity.
We refer to monomers with $\sus{1} > \sus{2}$ as field-aligning (FA) and monomers with $\sus{2} > \sus{1}$ as field-disaligning (FD)~\footnote{In past work~\cite{grasinger2020statistical,grasinger2020architected,cohen2016electroelasticity,cohen2016electromechanical}, field-aligning and field-disaligning monomers have alternatively been referred to as uniaxial and transversely isotropic monomers, respectively.}.

The energy of a single monomer has two contributions: the energy associated with separating charges and the electric potential of a dipole in an electric field~\cite{grasinger2020statistical}~\footnote{Where $\dipoleSusceptibility^{-1}$ denotes the generalized inverse of $\dipoleSusceptibility$.}:
\begin{equation}  \label{eq:monomer-energy}
    \um = \frac{1}{2\vperm} \dipole \cdot \dipoleSusceptibility^{-1} \dipole - \dipole \cdot \efield = \frac{\vperm \dsus}{2} \left(\efield \cdot \nvec\right)^2 - \frac{\vperm \sus{2}}{2} \emag^2,
\end{equation}
where $\dsus = \sus{2} - \sus{1}$.
We assume the monomer energy is such that $\um \sim \kB \T$ and that the energy of dipole-dipole interactions are much less than the thermal energy, i.e. $\vperm \left[\emag \times \max \left(\left\{\sus{1}, \sus{2}\right\}\right) \right]^2 / \mlen^3 \ll \kB \T$~\footnote{Together these assumptions imply that $\max \left(\left\{\susceptibility_\parallel, \susceptibility_\perp\right\}\right) / \mlen^3 \ll 1$}.
The latter is therefore neglected.

\newcommand{\ndv}{\hat{\bfv}}
Next the free energy corresponding to a constant temperature and constant electric field ensemble~\footnote{The minimum principle related to this free energy and its relation to the Helmholtz free energy can be found in \S 3.3  of~\cite{grasinger2020statistical}} of a DE chain is derived.
To this end, we derive the density of monomers oriented in the direction $\ndv$:
\begin{equation} \label{eq:density}
\density\left(\ndv\right) = \C \exp\left[-\unodim \left(\edir \cdot \ndv\right)^2 + \mults \cdot \ndv\right]
\end{equation}
where $\edir = \efield / \emag$, $\unodim = \vperm \emag^2 \dsus / 2 \kB \T$, and the unknowns, $\C$ and $\mults$, are determined by enforcing the end-to-end vector and normalization constraints
\begin{equation}
\label{eq:ccn} \N = \intoverSns{\density\left(\ndv\right)}, \quad \frac{\rvec}{\mlen} = \intoverSns{\density\left(\ndv\right) \ndv}, 
\end{equation}
and where $\unitsphere$ denotes the unit sphere.
In terms of $\density$, one can show that the free energy is approximately~\cite{grasinger2020statistical}:
\begin{equation} \label{eq:A-approx}
\A \approx \intoverSns{\left\lbrace \density \um + \kB \T \density \ln \density\right\rbrace} - \N \kB \T \ln \N.
\end{equation}

\newcommand{\Am}{\tilde{\A}_m}
\newcommand{\Ae}{\tilde{\A}_e}
\newcommand{\Ao}{\tilde{\A}_o}
There are still two challenges remaining in solving for $\C$ and $\mults$: \begin{inparaenum}[1)] \item the integrals in \eqref{eq:ccn} are difficult to evaluate and \item the resulting systems of equations are nonlinear. \end{inparaenum}
However, there are two limits in which a solution is tenable.
Let $\stch = \rmag / \N \mlen$ denote the absolute chain stretch (where $\rmag = |\rvec|$ and $\rdir = \rvec / \rmag$).
We derive approximate solutions in the limit of small stretch (i.e. $\stch \rightarrow 0$), which we denote as $\Asm$, and near the fully stretched limit (i.e. $\stch \rightarrow 1$), which we denote as $\Akg$.
In either case, the simplicity of the limit results from the observation that the Boltzmann term in the exponential of \eqref{eq:density} is invariant with respect to $\ndv \rightarrow -\ndv$.
Because of this symmetry, the kinematic constraint in \eqref{eq:ccn} can only be satisfied if $\left|\mults\right| \rightarrow 0$ as $\stch \rightarrow 0$~\footnote{A caveat: while this is certainly true when $\stch = 0$, it is only true that $\left|\mults\right| \ll 1$ in the neighborhood of $\stch = 0$ when $\mults$ is continuous (i.e. a phase transition is not occurring) at $\stch = 0$. Numerical solutions suggest this is a valid assumption~\cite{grasinger2020statistical}.} and, since $\density \rightarrow \delta\left(\rdir - \ndv\right)$ as $\stch \rightarrow 1$, this implies that $\mults$ diverges in the direction of the end-to-end vector (i.e. $\mults \rightarrow \infty \rdir$). 
In the former limit, we can use a Taylor expansion of $\density$ about $|\mults| = 0$ (\Fref{sec:short-comings}.B).
In the latter limit, we can neglect the Boltzmann factor (\Fref{app:fully-stretched}).
Then, using what is known about the limiting behavior, we construct a free energy approximation:
\begin{equation} \label{eq:asymptotic-A}
\begin{split}
  \A &= \Akg + \left(1 - \stch^2\right)\left(\lim_{\stch\rightarrow 0} \Asm - \lim_{\stch\rightarrow0} \Akg\right)\\
  &= \N \kB \T \left(\Am\left(\stch\right) + \Ae\left(\stch, \efield\right) + \Ao\left(\stch, \efield, \edir \cdot \rdir\right)\right)\,,
\end{split}
\end{equation}
where
\begin{equation}
\begin{split}
  \Am &= \stch \zmultzero + \ln \left(\frac{\zmultzero \csch \left[\zmultzero\right]}{4 \pi}\right) \\
  \Ae &= \frac{\stch \unodim}{\zmultzero} - \uOnodim + \\
      & \quad \left(1 - \stch^2 \right) \left[-\frac{\unodim}{3} + \ln \left(\frac{2 \sqrt{\unodim}}{\sqrt{\pi} \erfw}\right)\right] \\
  \Ao &= \unodim \left(1 - \frac{3 \stch}{\zmultzero}\right) \left(\edir \cdot \rdir\right)^2 
\end{split}
\end{equation}
are the (dimensionless) mechanical, electromechanical, and orientational contributions to the free energy, respectively; and where $\uOnodim = \vperm \emag^2 \sus{2} / 2 \kB \T$ and $\Langinv$ is the inverse Langevin function.
Note that, by construction, \eqref{eq:asymptotic-A} recovers the exact solution when $\unodim = 0$~\cite{kuhn1942beziehungen} and is exact in the limits of zero stretch and full stretch.
This approximation has been shown to agree well with numerical solutions in general~\cite{grasinger2020statistical}.

Given the free energy, one can obtain the net chain dipole, $\chainpolar$, by differentiating with respect to the electric field:
\begin{equation} \label{eq:chain-polarization}
\chainpolar = -\frac{\partial \A}{\partial \efield} = \intoverSns{\dipole\left(\ndv\right) \rho\left(\ndv\right)},
\end{equation}
which, as can be seen from \eqref{eq:asymptotic-A}, will depend on $\rvec$.
This dependence of the net chain dipole on the end-to-end vector is necessary to achieve a continuum-scale theory with a deformation dependent and anisotropic susceptibility.

Another feature of the free energy-stretch relationship of DE chains worth noting is that regardless of the sign or magnitude of $\unodim$, or the direction of chain stretch, numerical experiments suggest that the $\A / \kB \T$ vs $\stch$ curve is convex and its minimum is at zero stretch~\cite{grasinger2020statistical}.
This feature--namely the convexity and free energy minimum at zero stretch--has important physical implications.
It means that an individual chain will not stretch due to electrical excitation alone, or, equivalently, chains cannot support loads in compression.
Physically, this can be understood as a consequence of: \begin{inparaenum}[1)] \item the symmetry of \eqref{eq:monomer-energy} and \item the neglect of excluded volume effects. \end{inparaenum}
Indeed, if a monomer's direction is reversed its electrostatic energy does not change.
Since there is no energy penalty associated with large or small bond angles between neighboring monomers, and excluded volume effects are not taken into account, the chain is free to fold back on itself.
So in terms of the Boltzmann factor, a longer end-to-end vector is never any more favorable than a shorter end-to-end vector.
However, in terms of entropy, the shorter end-to-end vector is more favorable because there are a larger number of microstates that make up a shorter end-to-end vector than a longer end-to-end vector.
For these reasons, it can be argued, the free energy versus stretch relationship for a DE chain should be convex with its minimum at zero stretch.
And, in instances where this model is a good approximation of the physical system of interest, we should not expect that electrically-induced deformations occur as a result of individual chains stretching under electrical excitation alone.
Instead, as outlined previously and to be elaborated on further in Sections \ref{sec:orientational}-\ref{sec:shear}, \emph{intrinsic electrically-induced deformations are phenomena which can only be understood in the context of a cross-linked polymer network}.

It is likely, however, that when the bending stiffness (i.e. energy of rotating monomers about neighboring bonds) or excluded volume effects are nonnegligible, that the polymer chain has a free energy (local) minimum at finite stretch (and perhaps has multiple minima).
Similarly, chains consisting of monomers with fixed magnitude dipoles (i.e. ``frozen in'' dipoles) have an electrically-induced compressive stiffness~\cite{grasingerIPflexoelectricity}.
These cases could give rise to interesting behavior such as microbuckling~\cite{lakes1993microbuckling}, bistability, and phase transitions when loaded in compression~\cite{grekas2019cells,sun2020fibrous}.
While not relevant to the current study, the implications of these effects on electroactive polymer networks and electromechanical actuation present an interesting opportunity for future research.

\section{Short-comings of Linear Approximations: Gaussian-chain and Linearized Constraint} \label{sec:short-comings}
Before moving on to the next section in which we explore the continuum-scale implications of the chain-scale results, we show that neither \begin{inparaenum}[1)]
\item a Gaussian-chain approximation~\cite{warner2007liquid}--that is, modeling the chain as a random walk biased by the electric field--nor 
\item enforcing the end-to-end vector constraint to linear order,
\end{inparaenum}
properly captures the strong coupling between the chain end-to-end vector and net dipole of the chain.

\newcommand{\pr}{p}
\newcommand{\slen}{\boldsymbol{\ell}}
\newcommand{\bperp}{\unodim_{\perp}}
\newcommand{\bpara}{\unodim_{\parallel}}

\subsection{Gaussian-chain}
The Gaussian-chain approximation is as follows: the probability of a chain having an end-to-end vector $\rvec$ is taken to be proportional to the anisotropic Gaussian distribution:
\begin{equation} \label{eq:gaussian}
  \pr\left(\rvec\right) \propto \left(\det \slen\right)^{-1/2} \exp \left(-\frac{3}{2 \N \mlen} \rvec \cdot \slen^{-1} \rvec\right)
\end{equation}
where, in the context of liquid crystal elastomers, $\slen$ is referred to as the step length tensor~\cite{warner2007liquid}.
Equation \eqref{eq:gaussian} is the probability of a biased random walk of $\N$ steps with step length $\mlen$ starting at the origin and ending at $\rvec$, where the information about the biasing is encoded in $\slen$.
For dielectric elastomers, we would base the biasing of the random walk on the Boltzmann factor for a monomer with orientation $\nvec$.
Specifically, let $\mathbf{e}_3 = \edir$.
Then
\begin{equation} \label{eq:step-length}
  \slen = \mathbf{I} + \diag \left(f\left(\bperp\right), f\left(\bperp\right), f\left(\bpara\right) \right),
\end{equation}
where $\bpara \coloneqq \emag^2 \sus{1} / 2 \kB \T$, $\bperp \coloneqq \emag^2 \sus{2} / 2 \kB \T$, and $f$ is a monotonically increasing function of its argument that satisfies $f\left(0\right) = 0$.
Note, this is the most general definition of $\slen$ for dielectric elastomers which recovers the isotropic nature of the chain statistics when $\emag = 0$ and which is also consistent with the Boltzmann factor.
The next thing that one typically does is to say that the free energy of a chain is given by~\cite{warner2007liquid}:
\begin{equation} \label{eq:warner-free-energy}
  \A\left(\rvec; \efield\right) = -\kB \T \log \pr\left(\rvec; \efield\right).
\end{equation}
From here, however, it is not clear how to obtain the correct net chain dipole.
Consider \eqref{eq:chain-polarization} for example.
In this model, it is no longer the case that $-\partial \A / \partial \efield = \intoverSns{\dipole\left(\ndv\right) \rho\left(\ndv\right)}$.
If we wanted to be perfectly consistent with the notion of a biased random walk, then we would model the probability of a monomer being oriented in direction $\ndv$ as being given by $\density\left(\ndv\right) \propto \exp\left[-\unodim \left(\edir \cdot \ndv\right)^2\right]$; and let $\chainpolar = \intoverSns{\dipole\left(\ndv\right) \rho\left(\ndv\right)}$.
However, this would lead to $\chainpolar$ being completely independent of $\rvec$.
If instead we choose to model the net chain dipole as $\chainpolar = -\partial \A / \partial \efield$, then $\chainpolar$ depends on $\rvec$, but, it is easy to see from \eqref{eq:gaussian}, \eqref{eq:step-length}, and \eqref{eq:warner-free-energy}, that the net chain dipole cannot be misaligned with the electric field.
Hence, there can be no torque at the chain scale nor volumetric torque at the continuum-scale.
So not only is it unclear as to which expression would be more thermodynamically consistent (and less ad hoc), but, regardless of which choice one makes, the strong coupling between the end-to-end vector and net chain dipole cannot be captured through the Gaussian-chain approximation. 

\subsection{\label{sec:small-stretch} Linearized Constraint} 
The linearized constraint approximation is a good approximation as $\stch \rightarrow 0$ and is exact in this limit (see \S 6 of~\cite{grasinger2020statistical}).
We can obtain an approximate solution to the unknown multipliers, $\C$ and $\mults$ by a Taylor expansion of $\density$ about $\left|\mults\right| = 0$ to obtain:
\begin{equation} \label{eq:density-small-lambda}
  \density\left(\ndv\right) \approx \left(1 + \mults \cdot \ndv\right) \exp\left[-\unodim \left(\edir \cdot \ndv\right)^2\right]
\end{equation}
Substituting \eqref{eq:density-small-lambda} into the constraint equations, namely \eqref{eq:ccn}, and integrating gives~\footnote{
    When $\unodim$ is negative, we use:
    \begin{equation*}
        \erfw / \sqrt{\unodim} = \erf\left(\im \sqrt{|\unodim|}\right) / \im \sqrt{|\unodim|} = \erfi\left(\sqrt{|\unodim|}\right) / \sqrt{|\unodim|}
    \end{equation*}
    where $\erfi$ is the imaginary error function.
}
\begin{align}
    \label{eq:eap-sl-cn}
    \N &= 2 \pi^{3/2} \C \erfw / \sqrt{\uslnodim} \\
    \label{eq:eap-sl-crz}
    \frac{\rz}{\mlen} &= \pi \C \zmult \left(\frac{\sqrt{\pi} \erfw}{\uslnodim^{3/2}} - \frac{2 e^{-\uslnodim}}{\uslnodim}\right) \\ 
    \label{eq:eap-sl-crx}
    \frac{\rx}{\mlen} &= \pi \C \xmult \left(\frac{\sqrt{\pi} \left(2\uslnodim-1\right) \erfw}{2 \uslnodim^{3/2}} + \frac{ e^{-\uslnodim}}{\uslnodim}\right)
\end{align}
where $\mults = \left\{\alpha, 0, \lambda\right\}^T$.
Notice that \eqref{eq:eap-sl-cn}--\eqref{eq:eap-sl-crx} are linear in the unknowns and can be readily solved to obtain:
\begin{align} \label{eq:eap-sl-c}
    \C &= \Csl \\ \label{eq:eap-sl-z}
    \zmult &= \zmultsl \\ \label{eq:eap-sl-x}
    \xmult &= \xmultsl 
\end{align}
where $\stchz = \rz / \N \mlen$ and $\stchx = \rx / \N \mlen$.
Substituting \eqref{eq:eap-sl-c}, \eqref{eq:eap-sl-z}, and \eqref{eq:eap-sl-x} into \eqref{eq:density-small-lambda} and, subsequently, \eqref{eq:chain-polarization}, leads to a $\chainpolar$ which is independent of $\rvec$.
Hence, this approximation cannot capture either the extra stress contribution or an anisotropic polarization response.

\section{Orientational Chain Averaging and Continuum-scale Electroelasticity} \label{sec:orientational}
In order to relate the continuum mechanics of a DE to its chain-scale mechanics, we use the affine deformation assumption~\cite{treloar1975physics} and chain averaging.
At each material point in the stress-free, $\efield$-free state, it is assumed that \begin{inparaenum}[1)] \item chains have their most probable length, $\mlen \sqrt{\N}$, and \item since we expect isotropy in the absence of stress, the chain orientations are uniformly distributed. \end{inparaenum}
By the affine deformation assumption, each chain gets mapped from the reference configuration to the current configuration by the deformation gradient, $\F$, i.e. $\rvec = \F \Rvec$.
(A brief discussion of other polymer network models, their associated kinematic assumptions, and important physical considerations is given in \Fref{app:kinematic}.)
Finally, the free energy density is taken to be the product of the average chain free energy and the number of chains per unit volume, $\chaindensity$:
\begin{equation} \label{eq:energy-density}
\begin{split}
  \W\left(\F, \efield\right) &= \chaindensity \avg{\A\left(\F \Rvec, \efield\right)} \\&= \chaindensity \int \df \Rvec \; \left(\frac{\dirac\left(|\Rvec| - \mlen \sqrt{\N}\right)}{4 \pi}\right) \A\left(\F \Rvec, \efield\right),
\end{split}
\end{equation}
where $\chaindensity$ is the number of chains per unit volume and $\avg{\cdot}$ denotes an average over the distribution of chains~\footnote{Here we use $\W$ instead of $\mathcal{W}$ to denote the free energy density in order to emphasize that this is not the Helmholtz free energy density but its Legendre transform with respect to polarization.}.
Similarly, the polarization is given by
\begin{equation} \label{eq:polarization}
  \polarization\left(\F, \efield\right) = \chaindensity \avg{\chainpolar\left(\F \Rvec, \efield\right)} = -\frac{\partial \W}{\partial \efield},
\end{equation}
i.e. the product of the average net chain dipole and the number of chains per unit volume.

\section{Intrinsic Electrostriction}
\label{sec:electrostriction}

\begin{figure}
	\includegraphics[clip, trim=1.0cm 1.0cm 1.0cm 1.0cm, width=\linewidth]{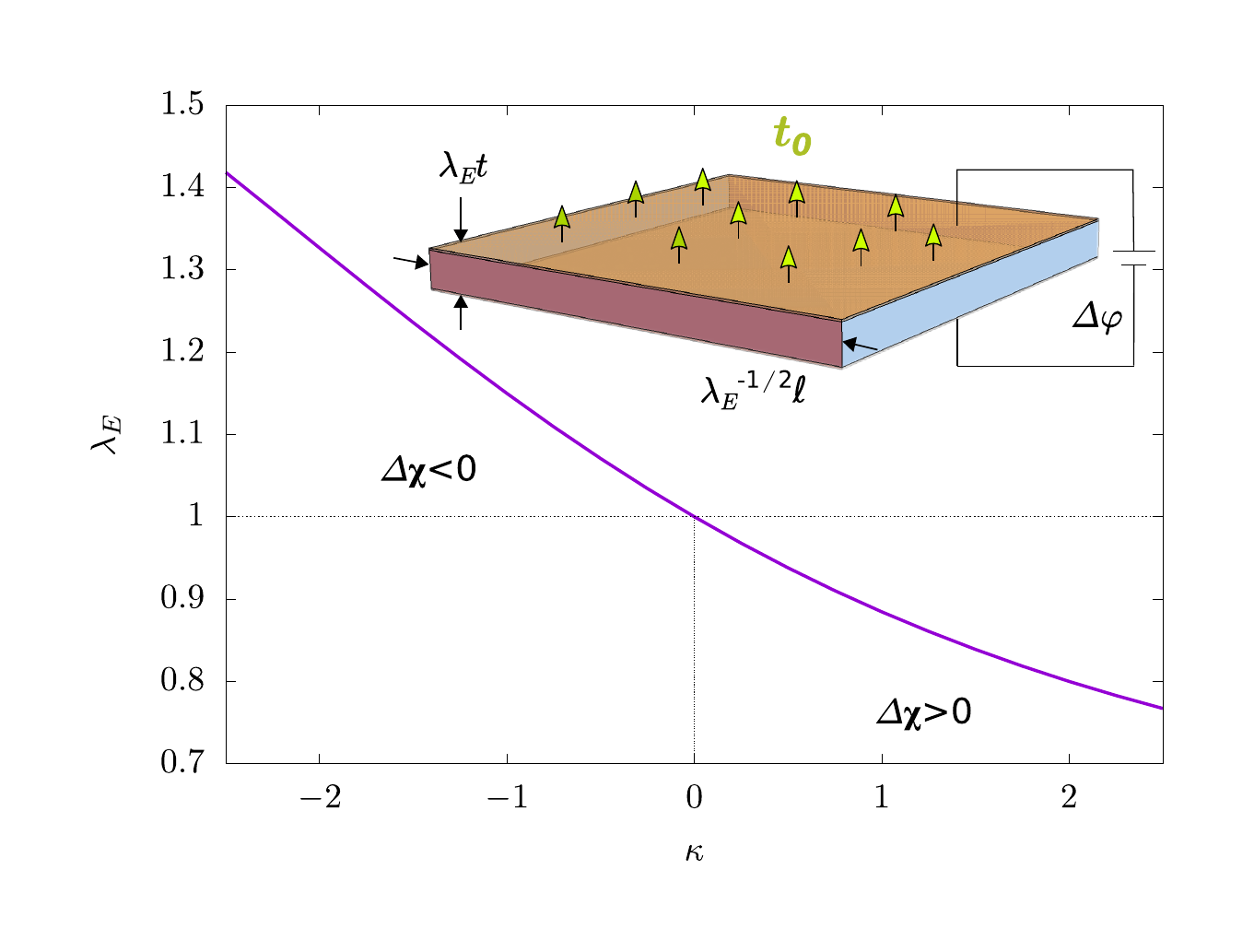}
  \caption{Electrostriction of a DEA with a fixed bottom surface and an applied traction, $\tvec$, to the top surface of the actuator that is equal and opposite to the Coulomb attraction (inset). Stretch across the thickness, $\pstche$, as a function of $\unodim = \emag^2 \dsus / 2 \kB \T$.}
	 \label{fig:electrostriction}
\end{figure}

Given the non-Gaussian model described above, we now turn to finding the electrostriction predicted by the model.
In order to isolate the phenomena of interest, we consider a typical DEA, but with the pressure from the electrodes removed.
This could be done by applying a traction, $\tvec$, to the outside surfaces of the electrodes that are equal and opposite to the Coulomb attraction and by ensuring the voltage difference is adjusted to keep the electric field constant when the distance between the electrodes changes (see \Fref{app:intrinsic-bcs}).
This is shown in \Fref{fig:electrostriction} (inset)~\footnote{See also \S 2.5 of ~\cite{grasinger2020architected}}.

In this case, the change in energy stored in the electric field (or equivalently, the work of the battery) and the work of the applied traction will cancel each other out; as a result, the equilibrium configuration will be the one that minimizes the ($\text{const } \T, \text{const } \efield$) free energy of the DE film.
To a good approximation, $\efield = \text{const} \times \ethree$ because $\deth \ll \delen$.
By symmetry, we assume the DE body undergoes homogeneous deformation of the form:
$\F = \diag \left(1 / \sqrt{\pstche}, 1 / \sqrt{\pstche}, \pstche\right)$.
Since the deformation and electric field are homogeneous, the free energy minimization reduces to a local problem, i.e., a (nonlinear) algebraic equation rather than a differential equation.
We use an adaptive Gauss-Kronrod quadrature for the integration and Newton's method for numerical optimization~\cite{Mathematica}.
The $\pstche$ which minimizes the free energy density for various $\unodim$ is shown in \Fref{fig:electrostriction}.
The model predicts an electrically-induced deformation of the DE film, even in the absence of pressure from the top and bottom electrodes.
When $\unodim  > 0$, that is, when the chain is made up of field-disaligning (FD) monomers, the film is compressed in the direction of $\edir$ (i.e. $\pstche < 1$).
Alternatively, when $\unodim < 0$ (chain consists of field-aligning (FA) monomers), the film actually elongates in the $\edir$ direction ($\pstche > 1$).

\newcommand{\pstchApprox}{\pstche^{\text{approx}}}
\newcommand{\auxf}{w_f^{*}}
We can obtain a closed-form approximation to the electrostrictive deformation by first taking a Taylor expansion of the inverse Langevin function about zero stretch and find
\begin{equation}
  \stch / \Langinvs = \frac{1}{3} - \frac{1}{5} \frac{\stchr^2}{\N} + \order{\stch^4}
\end{equation}
where $\stchr \coloneqq \rmag / \mlen \sqrt{\N}$ is the relative chain stretch~\footnote{Assuming chain stretch is small does not lead to inconsistencies for most cases of interest because chain stretch in the reference configuration scales like $1 / \sqrt{\N}$ (and $\N \sim 100$--$10$,$000$) so that, even when deformation is finite at the continuum-scale, the stretch of individual chains may still be considered small.}.
This provides an approximation for the free energy density:
\begin{widetext}
\begin{equation} \label{eq:Wstar-linearized}
\W\left(\F, \efield\right) 
= 
\chaindensity \kB \T \left( 
\N \left[ \auxf\left(\unodim\right) - \uOnodim\right] 
+ \left[\frac{3}{2} + \frac{2 \unodim}{15} - \auxf\left(\unodim\right)\right]\avg{\stchr^2} 
+ \frac{3 \unodim}{5} \avg{\stchr^2 \left(\edir \cdot \rdir\right)^2} 
+ \avg{\order{\stch^4 \N}}
\right)\,,
\end{equation}
\end{widetext}
where
\begin{equation*}
\auxf\left(\unodim\right) = \ln\left(\frac{2 \sqrt{\unodim}}{\sqrt{\pi}\erfw}\right).
\end{equation*}
Importantly, \emph{although using a linear approach for the statistical mechanics (e.g. Gaussian chain) misses the coupling between the chain end-to-end vector and net dipole, from \eqref{eq:Wstar-linearized} one can see that using a nonlinear statistical mechanics approach and then linearizing the nonliner result does still capture this coupling}; in other words, there are important differences between linearizing before ensemble averaging and linearizing after ensemble averaging.
Next the polymer network orientational averages in \eqref{eq:Wstar-linearized} can be obtained for any diagonal $\F$.
Indeed, let $\F = \diag \left(a_1, a_2, 1 / \left(a_1 a_2\right)\right)$.
Then:
\begin{equation} \label{eq:avgs-diag-F}
  \begin{split}
  \avg{\stchr^2} &= \frac{1}{3} \left(a_1^2 + a_2^2 + a_1^{-2} a_2^{-2}\right), \\
  \avg{\stchr^2 \left(\edir \cdot \rdir\right)^2} &= \frac{1}{3} a_1^{-2} a_2^{-2}.
  \end{split}
\end{equation}
For the intrinsic electrostriction outlined above $a_1 = a_2 = 1 / \sqrt{\pstche}$.
Substituting \eqref{eq:avgs-diag-F} into \eqref{eq:Wstar-linearized} and setting $\partial \W / \partial \pstche = 0$, we arrive at
\begin{equation}
  \pstchApprox = \left(\frac{4\unodim + 15\left(3 + \ln\pi\right) - 30\ln\left(\frac{2 \sqrt{\unodim}}{\erf\sqrt{\unodim}}\right)}{22\unodim + 15\left(3 + \ln\frac{\pi}{4}\right) - 30\ln\left(\frac{\sqrt{\unodim}}{\erf\sqrt{\unodim}}\right)}\right)^{1/3}
\end{equation}
which agrees well with \Fref{fig:electrostriction}.
The above approximation suggests that the electrostrictive deformation plateaus as $|\unodim| \rightarrow \infty$.
Indeed,
\begin{equation*}
  \lim_{\unodim \rightarrow \infty} \pstchApprox = \left(2 / 11\right)^{1/3}, \; \lim_{\unodim \rightarrow -\infty} = \left(13 / 2\right)^{1/3}.
\end{equation*}

Although explained at the continuum scale by a deformation-dependent susceptibility~\cite{zhao2008electrostriction,suo2010theory,cohen2016electromechanical,grasinger2020architected}, the nature of this electrostriction at the macromolecular scale has yet to be illustrated.
The physical reasoning is as follows: \begin{inparaenum}[1)] \item the tangent stiffness of a chain depends on its direction of stretch, \item chains experience an electrostatic torque (analogous to a dipole in an applied field)~\cite{grasinger2020statistical}, and \item elastomers are nearly incompressible~\footnote{
Elastomers are experimentally observed to be essentially incompressible~\cite{treloar1975physics}, and this is due to excluded volume effects that are difficult to consider in a statistical mechanics model.
Therefore, incompressibility is introduced \textit{a posteriori} as a postulate.
}. \end{inparaenum}
Regardless of its net dipole or the local electric field, every chain in the network has a minimum free energy length of zero, since we have neglected excluded volume effects for individual chains, and entropy is maximized in this configuration.
Therefore there is a positive stretching force in every chain, in all configurations, due to the incompressibility preventing the material collapsing to a point.
In the absence of electrical loading, all of the chains in the network have a similar stiffness and tension.
Further, due to incompressibility, shortening chains in one direction would force elongation of chains in other directions: this has a net energy increase.
However, under electrical stimulus, the isotropic symmetry of chain stiffnesses is broken such that a deformation which contracts some chains and stretches others leads to a net decrease in energy.
Specifically, FA chains are less stiff when aligned with the field direction and more stiff when orthogonal to the field~\cite{grasinger2020statistical}.
This, in part, explains the observed elongation in the field direction and contraction orthogonal to it when $\unodim < 0$ (vice versa for $\unodim > 0$).
The electrostatic torque on each DE chain in the network is a similar and related contribution.
For a chain consisting of FA monomers, for instance, some of its polarization is in the direction of its end-to-end vector (i.e. $\chainpolar \cdot \rvec \neq 0$).
As a result, there is a torque that is forcing the chain to align with $\spn \edir$.
It is clear that such rotations of the chains would lead to an elongation in the $\edir$ direction--which agrees with \Fref{fig:electrostriction}.
The reasoning is similar for chains consisting of FD monomers, but instead the torque forces these chains into the plane orthogonal to $\edir$ causing the film to compress in the $\edir$ direction.

Although the torque is nonzero for each individual chain, the average over the distribution vanishes (i.e. $\avg{\chainpolar \times \efield} = \mathbf{0}$).
However, in the next example, we will consider a case in which the average torque does not vanish and a volumetric torque is present, which gives rise to a shear-mode electromechanical coupling.

\section{Shear-mode Actuation}
\label{sec:shear}
\begin{figure}
	\includegraphics[clip, trim=1.0cm 1.0cm 1.0cm 1.0cm, width=\linewidth]{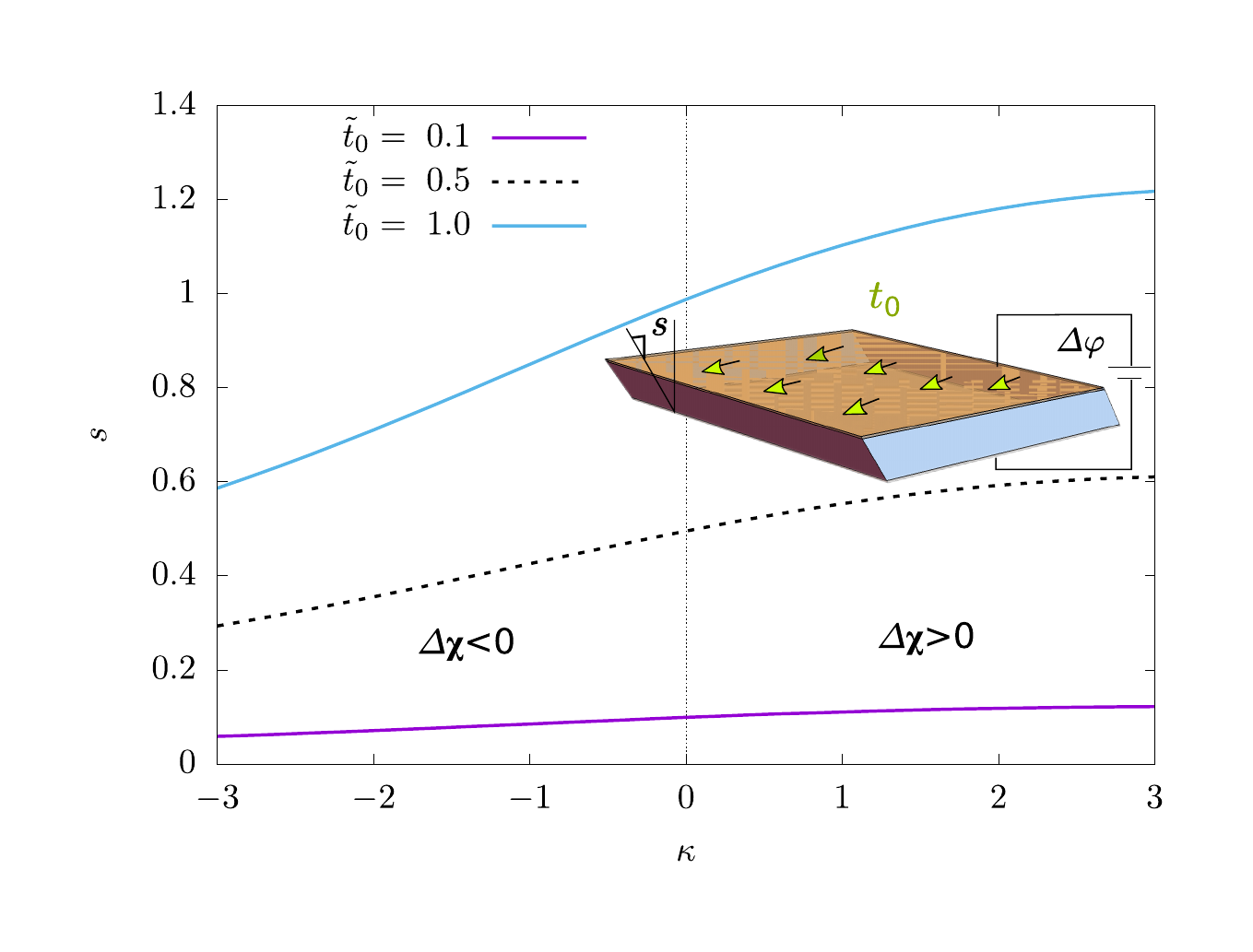}
  \caption{Shear-mode actuation of a DEA constrained to only undergo shear deformation (inset). 
  The film is prestressed, with $\tnodim = 0.1, 0.5$ and $1.0$, and then an electric field is applied. 
  Misalignment of the polarization with the field causes an electrostatic shear stress which increases the deformation for DEs with FD chains (i.e. $\emag^2 \dsus / 2 \kB \T = \unodim > 0$) and decreases the deformation for elastomers with FA chains (i.e. $\unodim < 0$). 
  The efficiency of the electrically induced deformation increases with $\tnodim$.
 }  
	\label{fig:shear}
\end{figure}

Again a thin DE film is considered with compliant electrodes on its top and bottom surfaces (e.g. \Fref{fig:DEA}.a).
However, here we assume the film is constrained such that it can only undergo homogeneous simple shear deformation~\footnote{This could be implemented by bonding the top surface of the DE film to an apparatus with rollers constraining its motion in the plane of shear}.
We also assume that the electric field is fixed and across the thickness of the film--which, neglecting fringe effects, would be realized by applying a voltage difference across the electrodes.
A schematic is shown in \Fref{fig:shear} (inset).

For simple shear, the deformation is of the form $\F = \mathbf{I} + \shearDef \eone \otimes \ethree$.
Let a traction, $\tvec = \left\{\begin{matrix}\tmag & 0 & 0\end{matrix} \right\}^T$, be applied to the top surface of the film.
The energy density of the electric field is neglected because it does not do work on the DE.
By homogeneity, the free energy is $\vol \times \left(\W - \tmag s\right)$ where $\vol$ is the volume of the body.

Let $\tnodim \coloneqq \tmag / \chaindensity \kB \T$ be the nondimensional applied traction.
Now consider the following experiment:
\begin{inparaenum}[1)]
\item the traction is applied and is held constant;
\item the traction causes initial shear deformation, $\shearDef_0$;
\item a voltage difference is applied across the electrodes such that there is electric field, $\efield$, in the DE;
\item for each $\efield$, a shear strain $\shearDef$ is observed.
\end{inparaenum}
The result, as predicted by the theory, is shown in \Fref{fig:shear}. 

As before with the electrostriction, we can obtain a closed-form approximation for the shear-mode deformation.
For simple shear:
\begin{equation} \label{eq:avgs-ss-F}
  \begin{split}
  \avg{\stchr^2} &= \frac{1}{3} \left(3 + \shearDef^2\right), \\
  \avg{\stchr^2 \left(\edir \cdot \rdir\right)^2} &= \frac{1}{3}.
  \end{split}
\end{equation}
Using \eqref{eq:avgs-ss-F} in \eqref{eq:Wstar-linearized} and solving for
$
  \partial \W / \partial \shearDef - \tmag = 0,
$
results in
\begin{equation} \label{eq:shear-deformation}
  \shearDef \approx \frac{\tnodim}{\frac{4}{45}\unodim + \frac{1}{3}\ln\frac{\pi}{4} - \frac{2}{3}\ln\left(\frac{\sqrt{\unodim}}{\erf\sqrt{\unodim}}\right)},
\end{equation}
and agrees well with \Fref{fig:shear}.

\Fref{fig:shear} and \eqref{eq:shear-deformation} show a shear electromechanical coupling.
It can be seen that DEs with chains consisting of FD monomers ($\dsus > 0$) spontaneously increase deformation with respect to an increasing electric field while DEs with FA ($\dsus < 0$) chains stiffen with increasing electric field such that the shear deformation decreases.
This can be understood through the affine deformation assumption.
Since chain end-to-end vectors are mapped under $\F$, the simple shear deformation serves to reduce average alignment with the field direction.
For chains with FD monomers, this causes their dipoles to align with the field--which is energentically favorable; whereas for chains with FA monomers this reduces average dipole alignment with the field, which is not energetically favorable.

More precisely, we can explain the shear-mode actuation in terms of volumeric torques, at the continuum-scale, and chain torque, at the macromolecular-scale.
Beginning with the former, consider again the total stress decomposition into mechanical and electrical parts.
While balance of angular momentum requires that the total stress tensor be symmetric, the individual contributions need not be.
To formulate the Maxwell stress, we approximate the polarization using the linearized form of $\W$ and \eqref{eq:polarization}.
The polarization, $\polarization$, is not aligned with $\edir$; indeed,
$
  \polarizationmag_1 = -2 \shearDef \vperm \dsus \emag / 5,
$
such that \emph{the Maxwell stress is asymmetric} for all finite $\shearDef$ and $\emag$.
This asymmetry causes an electrostatic volumetric torque within the material--which causes the top surface of the DE film to shear relative to the bottom.
While the Maxwell stress is asymmetric, the balance of angular momentum is satisfied by an equal and opposite asymmetry of the remaining contributions to the total stress.
Although here the volumetric torque leads to a shear-mode coupling, we mention that similar field induced volumetric torques can also be used for bending-couplings and shape-morphing~\cite{zhao2019mechanics,hajiesmaili2019reconfigurable,galipeau2013finite}.

In regards to the macromolecular-scale, just as in the case of electrostriction, there is an electrostatic torque on each of the individual chains in the network.
However in the present case, the average torque does not vanish; and further, it is equivalent to the skew-part of the Maxwell stress tensor.
Since the Maxwell stress is the sum of a rank one tensor and a symmetric tensor, we can obtain its skew-part by the Levi-Civita tensor, $\lct_{ijk}$:
\begin{equation}
\begin{split}
  \lct_{ijk} \left(\polarizationmag_i \emag_j + \frac{\vperm}{2} \emag_m \emag_m \delta_{ij}\right) \hat{\bfe}_{k} &= \avg{\chainpolar} \times \efield \\&= \avg{\chainpolar \times \efield}
\end{split}
\end{equation}
The average torque in the $\etwo$ direction is $2 \shearDef \vperm \dsus \emag^2 / 5$, which is nonvanishing for all finite $\shearDef$ and $\unodim$ and corresponds to $\maxwellij_{13}$.

\begin{figure}
	\includegraphics[clip, trim=1.0cm 1.0cm 1.0cm 0.0cm, width=\linewidth]{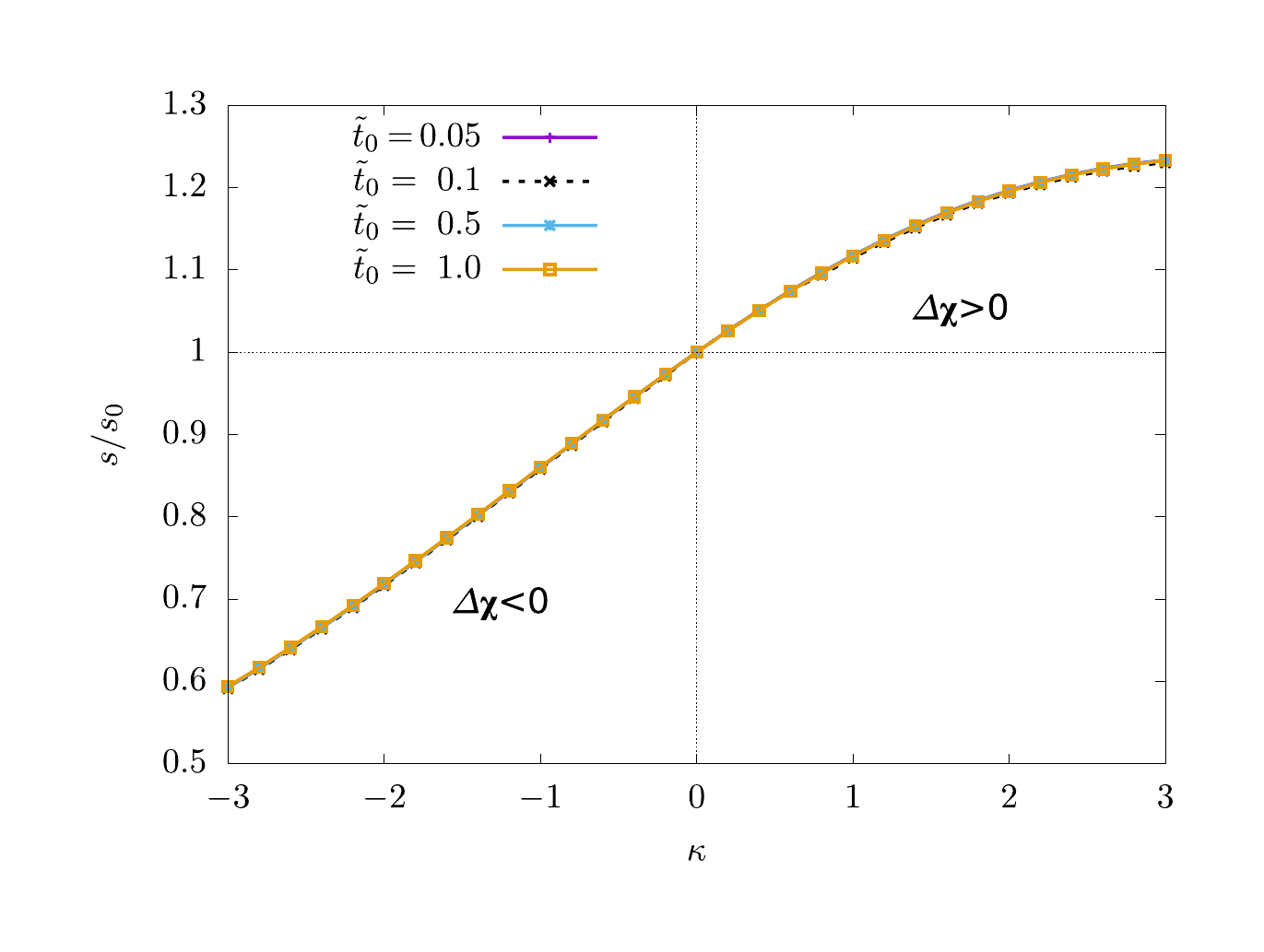}
	\caption{Shear-mode actuation of a DEA constrained to only undergo shear deformation.
   The electromechanical actuation (normalized by the initial deformation) is shown for applied tractions of $\tnodim = 0.1, 0.5$ and $1.0$.
  The curves collapse onto each other after normalization.
  Thus, the deformation appears to be proportional to the initial deformation, $\shearDef_0$.
}  
	\label{fig:shear-normalized}
\end{figure}

A notable feature of this shear-mode coupling is that it appears to scale (nearly exactly) linearly with the initial shear deformation, $\shearDef_0$.
Indeed, when normalizing the curves of $\shearDef$ in \Fref{fig:shear} by $\shearDef_0$, the curves collapse onto each other (\Fref{fig:shear-normalized}).
Similarly, the approximate solution given in \eqref{eq:shear-deformation} shows an exact linear scaling as $\shearDef \propto \tnodim$.
This indicates that, for a broad range of applied tractions, the strength of the electromechanical coupling can be tuned by adjusting the prestress in a straightforward way.
It should be noted, however, that beyond a certain shear deformation the assumption that the electric field is constant throughout the body and in the $\ethree$ direction will no longer be valid (i.e. geometric and fringe effects become nonnegligible). 

What we have shown is that the symmetry of DEs is broken when deformed.
The reason is that, while chain end-to-end vectors are isotropically distributed in the stress free network, they will, in general, not be isotropically distributed after deformation.
This breaking of symmetry means that the polarization is no longer restricted to be aligned with the local electric field and a broader class of electromechanical couplings are possible (e.g. bending, shape-morphing, etc.).

\section*{Conclusion}
In summary, we have presented a molecular-to-continuum-scale model of dielectric elastomers and used it to discover new electromechanical mechanisms and types of couplings: \begin{inparaenum}[1)] \item \emph{intrinsic} electrostriction; that is, a biaxial electromechanical coupling of a thin film DE actuator, despite the Coulomb attraction between the electrodes being counteracted; and, \item a shear electromechanical coupling where the electric field is orthogonal to the plane of shear. \end{inparaenum}
Each of these is a macroscale manifestation of a strong electromechanical coupling at the chain-scale, that cannot be captured through Gaussian chain approximations in statistical mechanics.

The Gaussian chain approximation is ubiquitous in the modeling of shape-responsive functional elastomers of various kinds, e.g. liquid crystal elastomers \cite{warner2007liquid}.
Therefore, a natural direction for future exploration is to identify other material systems and regimes in which one might expect to observe analogous unexpected couplings.

\begin{acknowledgments}
    This paper draws from the doctoral dissertation of Matthew Grasinger at Carnegie Mellon University \cite{grasinger2019multiscale}.
    We thank Gal deBotton, Timothy Breitzman, Pedro Ponte Casta\~{n}eda, and Pradeep Sharma for useful discussions.
    We acknowledge financial support from NSF (1635407), ARO (W911NF-17-1-0084), ONR (N00014-18-1-2528, N00014-18-1-2856), BSF (2018183), and AFOSR (MURI FA9550-18-1-0095).
    We acknowledge NSF for computing resources provided by Pittsburgh Supercomputing Center.
\end{acknowledgments}


\section*{\centering Appendix}
\appendix

\section{Statistical Mechanics of a Dielectric Elastomer Chain in the Fully Stretched Limit}
\label{app:fully-stretched}

To satisfy the kinematic constraint when the chain is near the fully stretched state (i.e. in the limit of $\stch \rightarrow 1$), we require that $\density \rightarrow \delta\left(\rdir - \ndv\right)$ as $\stch \rightarrow 1$, which implies that $\mults \rightarrow \infty \rdir$.
This is a consequence the symmetry of the monomer electrostatic energy (see \eqref{eq:monomer-energy}).
In this case, the Boltzmann factor can be neglected and the solution for $\density$ is well-known~\cite{kuhn1942beziehungen,grasinger2020statistical}:
\begin{equation}
  \densitykg\left(\ndv\right) \coloneqq \frac{\N \Langinv\left(\stch\right)}{4 \pi \sinh \left(\Langinv\left(\stch\right)\right)} \exp \left[\Langinv\left(\stch\right) \rdir \cdot \ndv\right].
\end{equation}

Substituting $\densitykg$ into \eqref{eq:A-approx} results in a free energy approximation that is exact in the limit $\stch \rightarrow 1$:
\begin{equation}
\begin{split}
	\Akg = \N \kB \T \Bigg[&\uznodim - \uOnodim + \frac{\stch}{\zmultzero}\left(\uxnodim - 2\uznodim\right)\\ 
	&-\stch \Langinv\left(\stch\right) - \log\left(\frac{\Langinv\left(\stch\right)}{4 \pi \sinh \left(\Langinv\left(\stch\right)\right)}\right)\Bigg]
\end{split}
\end{equation}
where $\uxnodim \coloneqq \emag_1^2 \dsus / 2 \kB \T$ and $\uznodim \coloneqq \emag_3^2 \dsus / 2 \kB \T$.

\section{Polymer Network Averaging Approaches and Kinematic Assumptions}
\label{app:kinematic}

Besides the affine deformation assumption, other polymer network models and associated kinematic assumptions exist in the literature.
The two most common kinematic assumptions in polymer network modelling are the affine deformation assumption and what we refer to as the \emph{cooperative network assumption}. 
The cooperative network assumption is typically used when a unit cell (i.e. representative volume element) consists of a finite number of chains.
In order to satisfy material frame indifference, the unit cell is rotated such that its axes are aligned with the principal frame and then the cell is stretched via the principal stretches~\cite{arruda1993threee,boyce2000constitutive}.
It is justified physically by ~\cite{arruda1993threee,boyce2000constitutive} as a way to model the ``cooperative behavior of the network''.
While this polymer network model and associated kinematic assumption have produced constitutive models that have been shown to agree well with experiments in standard rubber elasticity, it can produce seemingly nonphysical results in the context of field-responsive polymers.
We refer the reader to~\cite{cohen2018generalized}, and Chapters 1 and 3 of~\cite{grasinger2019multiscale}; in fact, in Ch. 3 of~\cite{grasinger2019multiscale}, one can see that the direction of the shear electromechanical actuation investigated in~\Fref{sec:shear} of this work is reversed (relative to the affine deformation assumption results) when the cooperative network assumption is used.
As a result, the experiment outlined in~\Fref{sec:shear}, or similar experiments, could be used as a means to indirectly probe the relationship between macroscopic-scale deformations and chain-scale deformations for various multifunctional elastomers of interest.

While there are also polymer network models which attempt to capture (non-affine) excluded volume effects~\cite{miehe2004micro}, these models generally include more fitting parameters and distract from the physics of interest.
We use the affine deformation assumption here because it relates the continuum scale deformation to the alignment and directionality of chains in the network in physically intuitive way.

\section{Boundary Conditions for Intrinsic Electrostriction Experiment} \label{app:intrinsic-bcs}

Let $\cbndy^{u}$ denote the upper electrode, $\cbndy^{b}$ denote the bottom electrode, and $\cbndy^{s}$ denote the four remaining sides of the DEA in the current configuration.
Then the precise boundary conditions for the intrinsic electrostriction experiment can be written as follows:
\begin{align*}
  \epot\left(\pos\right) &= \Delta \epot, \quad \bft\left(\pos\right) = \tvec = \frac{\vperm}{2} \left(1 + \sustens\left(\pstche\right) \right) \frac{\Delta \epot}{\pstche \deth} \hat{\bfe}_3 & \quad \forall \pos \in \cbndy^{u}; \\
\bft\left(\pos\right) &= 0 & \quad \forall \pos \in \cbndy^{s}; \\
\epot\left(\pos\right) &= 0, \quad \bfu\left(\pos\right) = 0 & \quad \forall \pos \in \cbndy^{b}.
\end{align*}

\bibliographystyle{apsrev4-1}
\bibliography{master}

\end{document}